\documentclass[12pt]{article}
\usepackage{amsmath}
\usepackage{bm}
\usepackage[dvipdfm]{graphicx}
\usepackage{ascmac}
\usepackage{mathrsfs}
\usepackage{multirow} % You have to install it.
\usepackage{feynmp} % feynman diagram
\usepackage{braket} % braket
\usepackage{layout}
\usepackage{cite}
\usepackage{booktabs}  % table
\newcommand{\vslash}{{v\hspace{-5.4pt}/}}
 \newcommand\pubnumber{}
 \newcommand\pubdate{}
\def\RCNP{Research Center for Nuclear Physics (RCNP), 
Osaka University, Ibaraki, Osaka, 567-0047, Japan}
\def\KEK{KEK Theory Center, Institute of Particle and Nuclear
Studies, High Energy Accelerator Research Organization, 1-1, Oho,
Ibaraki, 305-0801, Japan}

\def\Title#1{\begin{center} {\Large #1 } \end{center}}
\def\Author#1{\begin{center}{ \sc #1} \end{center}}
\def\Address#1{\begin{center}{ \it #1} \end{center}}

\newcommand\pubblock{\rightline{\begin{tabular}{l} \pubnumber\\
         \pubdate  \end{tabular}}}
\newenvironment{Abstract}{\begin{quotation}  }{\end{quotation}}
\newenvironment{Presented}{\begin{quotation} \begin{center} 
             PRESENTED AT\end{center}\bigskip 
      \begin{center}\begin{large}}{\end{large}\end{center} \end{quotation}}

\def\beq{\begin{equation}}
\def\eeq#1{\label{#1}\end{equation}}
\def\eeqn{\end{equation}}

\def\beqa{\begin{eqnarray}}
\def\eeqa#1{\label{#1}\end{eqnarray}}
\def\eeqan{\end{eqnarray}}

\let\bar=\overbar

\def\Dslash{\not{\hbox{\kern-4pt $D$}}}
\def\dslash{\not{\hbox{\kern-2pt $\del$}}}

\def\msb{{\bar{\ssstyle M \kern -1pt S}}}

\begin{document}
\begin{titlepage}
 \pubblock

\vfill
\Title{Exotic baryons from a heavy meson and a nucleon}
\vfill
% \Author{ Despina Reggiano\support}
\Author{Yasuhiro Yamaguchi, Shunsuke Ohkoda, and Atsushi Hosaka}
\Address{\RCNP}
\Author{Shigehiro Yasui}
\Address{\KEK} \vfill
% \Author{Yasuhiro}
% \Address{\napoli} \vfill
\begin{Abstract}
We evaluate a hadronic molecule formed by a heavy meson and a nucleon
respecting heavy quark symmetry.
The tensor force of $\pi$ exchange potential plays a dominate role to produce
an strong attraction in this system.
Solving coupled channel Schr\"odinger equations for $PN$ and $P^\ast N$,
we find many bound and resonant states with isospin $I=0$
 while there are few resonances in $I=1$ state.
The rich structures with $I=0$ indicate that the spectrum of
heavy baryons near the threshold is influenced by the contributions from
such hadron composite structures.

% I describe a case study of Mesmeric influence on a physiological  reaction
% in two  Albanian subjects.
\end{Abstract}
\vfill
\begin{Presented}
The 5th International Workshop on Charm Physics (Charm2012) \\
Honolulu, Hawaii, May 14--17  2012
% Symposium of the Physical Society of Rome\\
% Rome, Italy,  August 13--17, 1797
\end{Presented}
\vfill
\end{titlepage}
\def\thefootnote{\fnsymbol{footnote}}
\setcounter{footnote}{0}
%

%==============================================
\section{Introduction}
%==============================================
Hadronic molecular picture gives us new aspect for the structure of
exotic hadrons.
Some candidates of molecular state have been observed in
charmonium and bottomonium regions~\cite{Brambilla:2010cs}.
In particular, the twin $Z_b$ resonances which are new evidence of
hadronic composite in bottom region get much attention recently.
The $Z_b(10610)$ and $Z_b(10650)$ were reported by Belle
in 2011~\cite{Collaboration:2011gja,Belle:2011aa,Adachi:2012im}.
Since the $Z_b$'s have electric charge, they cannot be explained by
simple $b\bar{b}$ state.
Its quark content needs at least $b\bar{b}q\bar{q}$ where $q$ is $u$ or
$d$ quark and they have exotic structure deflected from standard hadrons.
Furthermore these masses are $M(Z_b(10610))=10607.2 \pm2.0$ MeV and
$M(Z_b(10650))=10652.2 \pm1.5$ MeV which are close to $B\bar{B}^\ast$
and $B^\ast\bar{B}^\ast$ thresholds.
Hence, they are considered as $B\bar{B}^\ast$ and $B^\ast\bar{B}^\ast$ molecules.
In the theoretical research also, some authors discuss the states as
meson-meson molecule~\cite{Voloshin:2011qa,Bondar:2011ev,Ohkoda:2011vj}.

In the studies of hadronic molecules with heavy meson such as the $Z_b$'s,
$\pi$ exchange interaction with heavy quark symmetry attracts attention
to design the hadronic composite structure~\cite{Tornqvist:2004qy,Cohen:2005bx,Ohkoda:2011vj,Yasui:2009bz,Yamaguchi:2011xb,Yamaguchi:2011qw,Ohkoda:2012hv}.
The heavy quark symmetry appears in the heavy quark mass limit
$(m_Q\rightarrow \infty)$~\cite{Isgur:1989vq,Isgur:1991wq}.
Because spin-spin interaction between quarks is suppressed under this
limit, heavy pseudoscalar meson $P$ and heavy vector meson $P^\ast$
are degenerate.
Indeed, mass splitting among $P$ and $P^\ast$ is very small, $BB^\ast$
mass splitting is about 45 MeV and $DD^\ast$ mass splitting is about 140
MeV.
% Compared to this, in strangeness sector, $KK^\ast$ mass splitting is so
% large, about 400 MeV.
This degeneracy provides $PP^\ast \pi$ vertex and $\pi$ exchange
potential in the system.
We note that $PP\pi$ vertex is forbidden due to parity conservation, and
$KK^\ast \pi$ vertex in strangeness sector is suppressed by large
$KK^\ast$ mass splitting, about 400 MeV.

The $\pi$ exchange potential is of great interest as one of the most
important meson exchange forces.
In particular, the tensor force of $\pi$ exchange plays a crucial role.
It is known that the tensor force mixing the states with different angular momentum, $L$ and
$L\pm 2$, yields a strong attraction between hadrons.
That is a leading mechanism of the binding of nuclei~\cite{Ericson_Weise}.
% The nature of $\pi$ exchange interaction is important in hadronic
% molecule.
% The long range force designs a loosely bound state as seen in
% deuteron.
In hadronic molecule with heavy meson, the $\pi$ exchange interaction
with $PP^\ast \pi$ vertex is expected to generate a strong attraction.

In this study, we investigate exotic and non-exotic heavy baryon states formed by a heavy meson and
a nucleon.
For exotic states, we consider $\bar{D}^{(\ast)}N$ and $B^{(\ast)}N$ molecule whose
minimal quark content is $\bar{Q}qqqq$ without $q\bar{q}$ annihilation,
where $Q$ denote heavy quark.
The non-exotic states are described as $D^{(\ast)}N$ and $\bar{B}^{(\ast)}N$ molecules.
They are constructed by $Q\bar{q}qqq$ and can be coupled with ordinary baryon
states like $\Lambda_{\text{c}}$, $\Sigma_{\text{c}}$ and so on.
Hence, we have to take into account many channel couplings in this states.
However, we leave this problem as feature work.
We focus on only $D^{(\ast)}N(\bar{B}^{(\ast)}N)$ channels.% with $\pi$ exchange.
As a meson exchange interaction, we consider not only $\pi$ exchange but
also vector meson $(\rho,\omega)$ exchange to estimate short range
interaction.
We study bound states and resonances by solving coupled channel equations
for $PN$ and $P^\ast N$.

% In heavy meson-nucleon molecule, $\pi$ exchange potential appears
% through channel coupling with $P^\ast N$.
% We analyze coupled channel equation for $PN$ and $P^\ast N$ channels.

%==============================================
\section{Interactions}
%==============================================
We investigate two-body states of $P^{(\ast)} N$ with
$J^P=1/2^\pm, 3/2^\pm, 5/2^\pm, 7/2^\pm$, where $J$ is
total angular momentum and $P$ is parity.
Moreover, these states take isospin $I=0$ or $1$.
Each states has three or four channels mixed by tensor force of $\pi$
exchange potential.
The channels for given $J^P$ are classified by orbital angular momentum $L$ and spin $S$
as summarized in Table~\ref{channels}.
The higher angular momentum states induced by tensor force are important
to generate attraction.
We consider full channel couplings of $PN$ and $P^\ast N$ channels.

% We focus on the $\pi$ exchange interaction between a heavy meson and a
% nucleon due to the degeneracy of $P$ and $P^\ast$.
% The two-body states of $P^{(\ast)}N$ are
% classified by isospin $I$, total angular momentum $J$, and parity $P$.
% For given $J^P$, various coupled channels are summarized in
% Table~\ref{channels}.
% The higher angular momentum states are important to introduce tensor
% force in the system.
% Therefore, we consider full channel couplings of $PN$ and $P^\ast N$.

%===========================================
\begin{table}[t]
\begin{center}
\caption{\small Various coupled channels for a 
given quantum number $J^P$.  }
\label{channels}
\vspace*{0.5cm}
{%\small 
\begin{tabular}{ c  | c c c c}
\hline
$J^P$ &  \multicolumn{4}{c }{channels} \\
\hline
$1/2^-$ &$PN(^2S_{1/2})$&$P^\ast N(^2S_{1/2})$&$P^\ast N(^4D_{1/2})$& \\
$1/2^+$ &$PN(^2P_{1/2})$&$P^\ast N(^2P_{1/2})$&$P^\ast N(^4P_{1/2})$& \\
$3/2^-$ & $PN(^2D_{3/2})$&$P^\ast N(^4S_{3/2})$&$P^\ast
	      N(^4D_{3/2})$&$P^\ast N(^2D_{3/2})$ \\
$3/2^+$ & $PN(^2P_{3/2})$&$P^\ast N(^2P_{3/2})$&$P^\ast
	      N(^4P_{3/2})$&$P^\ast N(^4F_{3/2})$ \\
$5/2^-$&$PN(^2D_{5/2})$&$P^\ast N(^2D_{5/2})$&$P^\ast
	      N(^4D_{5/2})$&$P^\ast N(^4G_{5/2})$ \\
$5/2^+$&$PN(^2F_{5/2})$&$P^\ast N(^4P_{5/2})$&$P^\ast
	      N(^2F_{5/2})$&$P^\ast N(^4F_{5/2})$ \\
$7/2^-$&$PN(^2G_{7/2})$&$P^\ast N(^4D_{7/2})$&$P^\ast
	      N(^2G_{7/2})$&$P^\ast N(^4G_{7/2})$ \\
$7/2^+$&$PN(^2F_{7/2})$&$P^\ast N(^2F_{7/2})$&$P^\ast
	      N(^4F_{7/2})$&$P^\ast N(^4H_{7/2})$ \\
\hline
\end{tabular}
}
\end{center}
\end{table}
%===========================================

The interaction Lagrangians for $\pi$ meson and vector meson ($\rho$ and $\omega$)
exchanges among $P^{(\ast)}N$ are described as satisfying the heavy
quark symmetry and chiral symmetry~\cite{Casalbuoni}.
They are written by
\begin{align}
{\cal L}_{\pi HH} &=   ig_\pi \mbox{Tr} \left[
H_b\gamma_\mu\gamma_5 A^\mu_{ba}\bar{H}_a \right]   \, , 
\label{LpiHH}\\
{\cal L}_{v HH} &=  -i\beta\mbox{Tr} \left[ H_b
v^\mu(\rho_\mu)_{ba}\bar{H}_a \right]
+i\lambda\mbox{Tr} \left[
H_b\sigma^{\mu\nu}F_{\mu\nu}(\rho)_{ba}\bar{H}_a \right] \, , 
\label{LvHH}
\end{align}
where the subscripts $\pi$ and $v$ are for the pion and vector meson
($\rho$ and $\omega$) interactions, and $v^\mu$ is the four-velocity of
a heavy quark. The heavy meson field $H$ is defined by 
\begin{align}
H_a   &= \frac{1+\vslash}{2}\left[P^\ast_{a\,\mu}\gamma^\mu-P_a\gamma_5\right] \, ,  \\
\bar H_a  &= \gamma_0 H^\dagger_a \gamma_0 \, ,
\end{align}
where the subscript $a$ is for light flavors, $u$, $d$.
% The pseudoscalar and vector fields are normalized as
% \begin{align}
%  \langle 0|P|P(p_\mu)\rangle&=\sqrt{p^0} \, , \\
%  \langle 0|P^{\ast}_\mu|P^{\ast}(p_\mu,\lambda)\rangle&=\epsilon(\lambda)_\mu\sqrt{p^0} \, ,
% \end{align}
% where $\epsilon(\lambda)_\mu$ is the polarization vector of $P^\ast$
% with polarization $\lambda$. 
The axial current of light
flavors is written by
\begin{align}
\displaystyle  A^\mu&=\frac{1}{2}\left(\xi^\dagger\partial^\mu
 \xi-\xi\partial^\mu \xi^\dagger\right) \, ,
\end{align}
with $\xi=\exp{(i\hat{\pi}/f_\pi)}$ and the pion decay
constant $f_\pi=132$ MeV. The pion field is defined by
\begin{align}
 \hat{\pi}&=\left(
\begin{array}{cc}
\displaystyle \frac{\pi^0}{\sqrt{2}}&\pi^+ \\
 \pi^- &\displaystyle -\frac{\pi^0}{\sqrt{2}}
\end{array}
\right).
\end{align}
For vector meson, its field is defined by
\begin{align}
 \rho^\mu&=i\frac{g_V}{\sqrt{2}}\hat{\rho}_\mu \, , \\
 \hat{\rho}_\mu&=\left(
\begin{array}{cc}
 \displaystyle \frac{\rho^0}{\sqrt{2}}&\rho^+ \\
 \rho^- &\displaystyle -\frac{\rho^0}{\sqrt{2}}
\end{array}
\right)_\mu \, ,
\end{align}
where $g_V$ is the gauge coupling constant of hidden local
symmetry~\cite{Bando:1987br}.
The vector meson field tensor is written by
$F_{\mu\nu}(\rho)=\partial_\mu \rho_\nu-\partial_\nu \rho_\mu+[\rho_\mu,\rho_\nu]$.
From Eqs.~\eqref{LpiHH} and \eqref{LvHH}, we obtain the pion and vector
meson vertices in the static approximation $v^\mu=(1,\vec{0})$.

The coupling constant $g_\pi$ for pion is fixed from the strong decay of
$D^\ast\rightarrow D\pi$~\cite{Yasui:2009bz}.
The coupling constants $\beta$ and $\lambda$ are determined by radiative
decays of $D^\ast$ and semileptonic decays of $B$~\cite{Isola:2001bn}.

For vertices of a meson and nucleons, we employ Bonn model~\cite{machleidt} as
\begin{align}
 {\cal L}_{\pi NN} &= \sqrt{2} ig_{\pi
NN}\bar{N}\gamma_5 \hat{\pi} N \, , \label{LpiNN} \\
{\cal L}_{vNN} &= \sqrt{2} g_{vNN}\left[\bar{N} \gamma_\mu \hat{\rho}^\mu N 
+\frac{\kappa}{2m_N}\bar{N} \sigma_{\mu\nu} \partial^\nu \hat{\rho}^\mu
N \right] \, , \label{LvNN}
\end{align}
where $N=(p,n)^T$ is the nucleon field.

To take into account internal structure of hadrons, we introduce form factors $F$
associated with finite size of the mesons and nucleons:
\begin{align}
 F_\alpha(\Lambda,\vec{q})&=\frac{\Lambda^2-m^2_\alpha}{\Lambda^2+|\vec{q}\,^2|}
\end{align}
where $m_\alpha$ and $\vec{q}$ are the mass and three-momentum of the
incoming meson $\alpha$ ($=\pi,\rho,\omega$).
The cutoff parameter $\Lambda=\Lambda_N$ for nucleon vertex is determined to reproduce the
properties of deuteron by solving $NN$ system.
In order to fix another cutoff $\Lambda=\Lambda_P$ for heavy meson vertex,
we assume the relation between a ratio of cutoff and a ratio of size of
hadron, written as $\Lambda_P/\Lambda_N=r_N/r_P$ where $r_N$ and $r_P$ are
the sizes of nucleon and heavy meson respectively.
The ratio of size is estimated from quark model
calculation~\cite{Yasui:2009bz} and
we obtain $\Lambda_D=1.35\Lambda_N$ for $D$ meson and $\Lambda_B=1.29\Lambda_N$ for $B$ meson. 
The results are shown in Table~\ref{cutoff}.

%===============================================================================
\begin{table}[htbp]
\begin{center}
\caption{The cutoff parameters for nucleon ($\Lambda_N$), $D$ meson
 ($\Lambda_D$) and $B$ meson ($\Lambda_B$) vertices.}
\label{cutoff}
\begin{tabular}{cccc}
\hline
Potential &  $\Lambda_N$ [MeV] & $\Lambda_D$ [MeV] & $\Lambda_B$ [MeV] \\
\hline
$\pi$  & 830  & 1121  &  1070  \\
$\pi,\rho,\omega$ & 846  &  1142  & 1091 \\ \hline
\end{tabular}
\end{center}
\end{table}
%===============================================================================

%==============================================
\section{Bound states and resonances in $\bar{D}N$ and $BN$}
%==============================================
In this section, let us show the results for analyzing bound and
resonant states given by solving the coupled channel Schr\"odinger
equations for $PN$ and $P^\ast N$. 

First, we discuss the results for $\bar{D}N$ and $BN$ states.
They are truly exotic state without $q\bar{q}$ annihilation.
For $I=0$ state, we found some bound states for $\bar{D}N$ and $BN$ in $J^p=1/2^-$.
The binding energies and relative distances for bound states
are summarized in Table~\ref{1/2bound}.
In this analysis, we compare results of two potentials.
One is only $\pi$ exchange potential and the other is $\pi\rho\omega$
exchange potential.
As shown in Table~\ref{1/2bound}, the results of $\pi$ exchange potential
are almost the same as those of $\pi\rho\,\omega$ exchange potential.
This result indicates that $\pi$ exchange plays a dominant role to
compose the bound state while vector meson exchanges are minor.
In particular, the tensor force causing mixing between $PN$ and $P^\ast N$
states with different angular momentum $\Delta L=2$ yields strong attraction.
For $BN$ state, small $BB^\ast$ mass splitting induces strong $BB^\ast$
mixing and tensor force.
Therefore, $BN$ state is more bound than $\bar{D}N$ state.

%%%%%%%%%%%%%%%%%%%%%%%%%%%%%%%%%%%%%%%%%%%%%%%%%%%%%%%%%%%%%%%%
\begin{table}[htbp]
 \begin{center}
  \caption{Binding energies $E_B$ and relative radii $\sqrt{\langle
  r^2\rangle}$ of $\bar{D}N$ and $BN$ bound states in
  $(I,J^P)=(0,1/2^-)$. Results for $\pi$ and $\pi\rho\,\omega$ potential
  are compared.}
  \label{1/2bound}
  \begin{tabular}{ccccc}
   \hline
   &$\bar{D}N(\pi)$ &$\bar{D}N(\pi\rho\,\omega)$ &$BN(\pi)$
   &$BN(\pi\rho\,\omega)$ \\
   \hline
   $E_B$ [MeV] &1.60 &2.13&19.50&23.04 \\
   $\sqrt{\langle r^2\rangle}$ [fm] &3.5&3.2&1.3&1.2 \\
   \hline
  \end{tabular}
 \end{center}
\end{table}
%%%%%%%%%%%%%%%%%%%%%%%%%%%%%%%%%%%%%%%%%%%%%%%%%%%%%%%%%%%%%%%%

The corresponding root mean square radii are over 3 fm for $\bar{D}N$
and over 1 fm for $BN$ respectively.
Both radii are larger than typical hadron size of order 1 fm, justifying
the hadronic composite structure of the present states.

\begin{figure}[htb]
\centering
\includegraphics[height=0.4\textwidth]{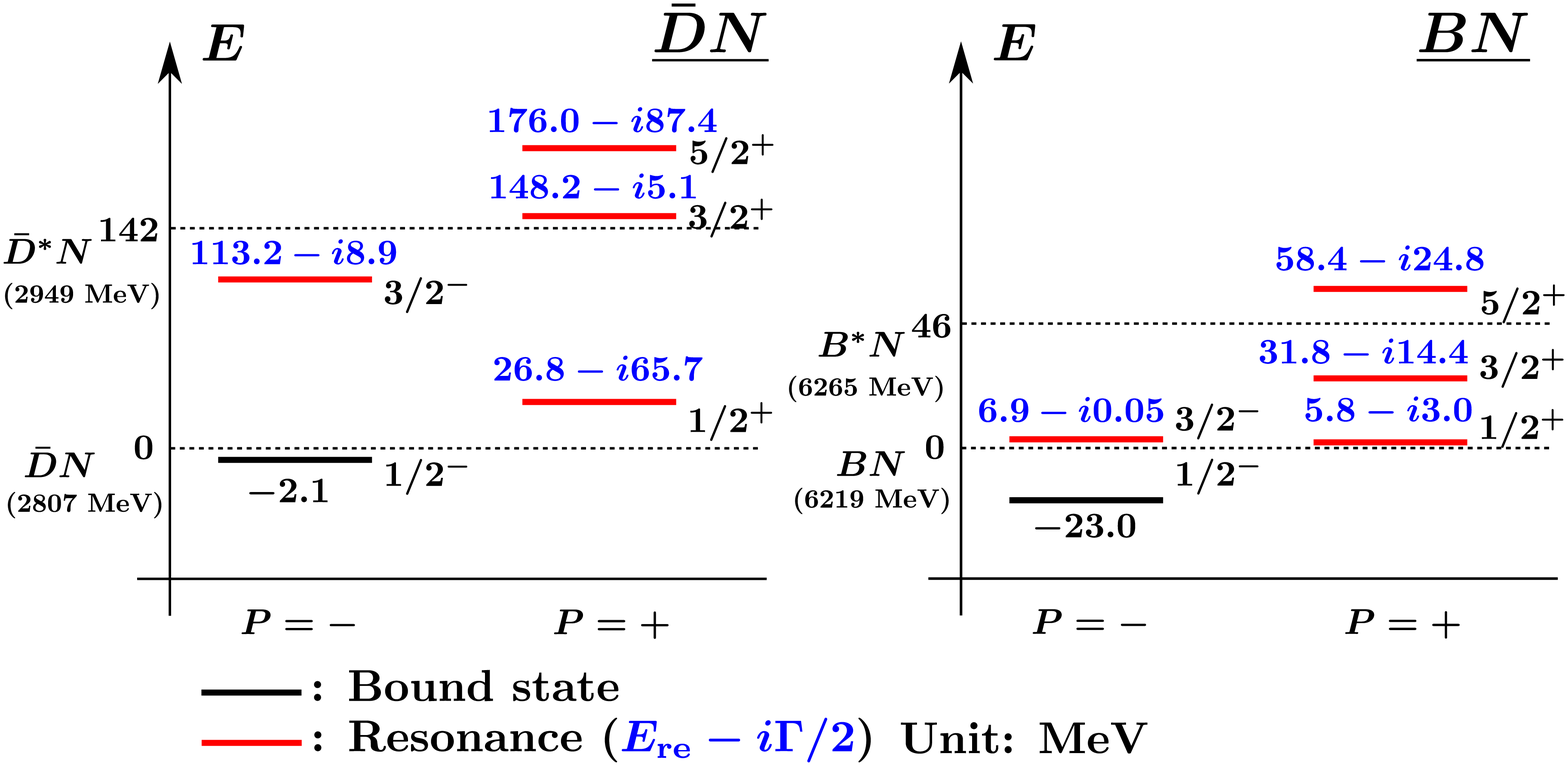}
 \caption{Bound states and resonances for $\bar{D}N$ and $BN$ with $I=0$.
 The energies are measured from the lowest
  thresholds ($\bar{D}N$ and $BN$). The binding energy is given as a
  real negative value, and the resonance energy $E_{\mathrm{re}}$ and
  decay width $\Gamma$ are given as $E_{\mathrm{re}}-i\Gamma/2$. The
  values are given when the $\pi\rho\,\omega$ potential is used.}
\label{Energy-LevelDbarN}
\end{figure}

In addition to bound states, we found some resonances near the threshold
in $J^P=1/2^+,3/2^\pm$ and $5/2^+$ states with $I=0$.
Energy levels of bound states and resonances are shown in
Fig.~\ref{Energy-LevelDbarN}.
The tensor force of $\pi$ exchange potential produces a strong
attraction and constructs various resonant states in $\bar{D}N$ and $BN$.
In the $J^P=3/2^-$ state, we find an interesting structure, that is
Feshbach resonance.
The mechanism of the resonance can be understood by the presence of
$P^\ast N$ quasi-bound state above $PN$ threshold.
Indeed, when the $PN(^2D_{3/2})$ channel is ignored and only
$P^\ast N(^4S_{3/2})$, $P^\ast N(^4D_{3/2})$ and $P^\ast N(^2D_{3/2})$
channels are considered, there are bound states of $\bar{D}^\ast N$ and
$B^\ast N$.

%%%%%%%%%%%%%%%%%%%%%%%%%%%%%%%%%%%%%%%%%%%%%%%%%%%%%%%%%%%%%%%%
%%%%%%%%%%%%%%%%%%%%%%%%%%%%%%%%%%%%%%%%%%%%%%%%%%%%%%%%%%%%%%%%

Finally, we found no bound state and
resonance in $I=1$ state.
The isotriplet state cannot obtain a sufficient attraction from the
tensor force of $\pi$ exchange due to small isospin factor
$\vec{\tau}_P\cdot\vec{\tau}_N=1$ while $\vec{\tau}_P\cdot\vec{\tau}_N$
is $-3$ with $I=0$.

%==============================================
\section{Bound states and resonances in $DN$ and $\bar{B}N$}
%==============================================
Let us now move to the study of non-exotic states ($DN$ and $\bar{B}N$).
The $DN$ state is calculated in analogy with $\bar{D}N$.
But $\pi$ exchange potential and $\omega$ exchange potential get
opposite sign due to $G$-parity transformation.

\begin{figure}[htbp]
\centering
\includegraphics[height=0.4\textwidth]{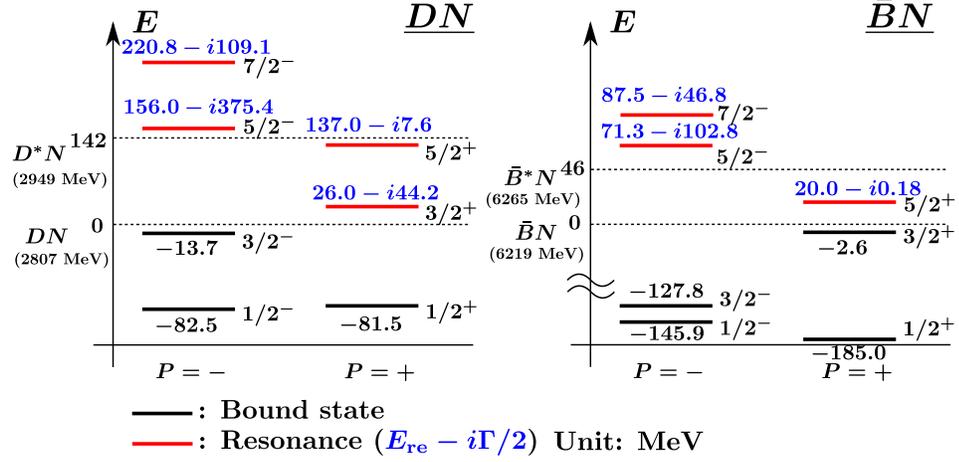}
 \caption{Bound states and resonances for $DN$ and $\bar{B}N$ with
 $I=0$.  The values are given when the $\pi\rho\,\omega$ potential is used.}
\label{Energy-LevelDN}
\end{figure}

As shown the energy-level in Fig.~\ref{Energy-LevelDN},
many bound and resonant states are present in $I=0$ state.
The bound states appear in $J^P=1/2^\pm$ and $3/2^-$ for $DN$ state, and in
$J^P=1/2^\pm$ and $3/2^\pm$ for $\bar{B}N$ state.
Above the $DN$ and $\bar{B}N$ thresholds, resonant states are found in 
$J^P=3/2^+,5/2^\pm$ and $7/2^-$ for $DN$ state, and in $J^P=5/2^\pm$ and
$7/2^-$ for $\bar{B}N$ state.
In comparison with $\bar{D}N$ and $BN$ states, non-exotic states are
more bound.
That strong attractive force comes from diagonal terms of $DN(\bar{B}N)$
potentials.
Although diagonal parts of $\bar{D}N$ potential behave repulsively, the
diagonal terms of $DN$ are attractively because $G$-parity
transformation switches signs of $\pi$ and $\omega$ exchanges.

In particular, for low angular momentum states, they compose deeply
bound states with small radius which is less than 1 fm.
Hence, constituent hadrons, heavy meson and nucleon, overlap each other
and they may construct a compact object.
%  which is not described as
% molecular state.
For such states, we have to consider further channel couplings with charmed
baryon state like $\Lambda_{\text{c}}$ and $\Sigma_{\text{c}}$.% $\pi \Sigma^{(\ast)}_{\text{c}}$.

%%%%%%%%%%%%%%%%%%%%%%%%%%%%%%%%%%%%%%%%%%%%%%%%%%%%%%%%%%%%%%%%
%%%%%%%%%%%%%%%%%%%%%%%%%%%%%%%%%%%%%%%%%%%%%%%%%%%%%%%%%%%%%%%%

%%%%%%%%%%%%%%%%%%%%%%%%%%%%%%%%%%%%%%%%%%%%%%%%%%%%%%%%%%%%%%%%
\begin{figure}[htbp]
\centering
\includegraphics[height=0.4\textwidth]{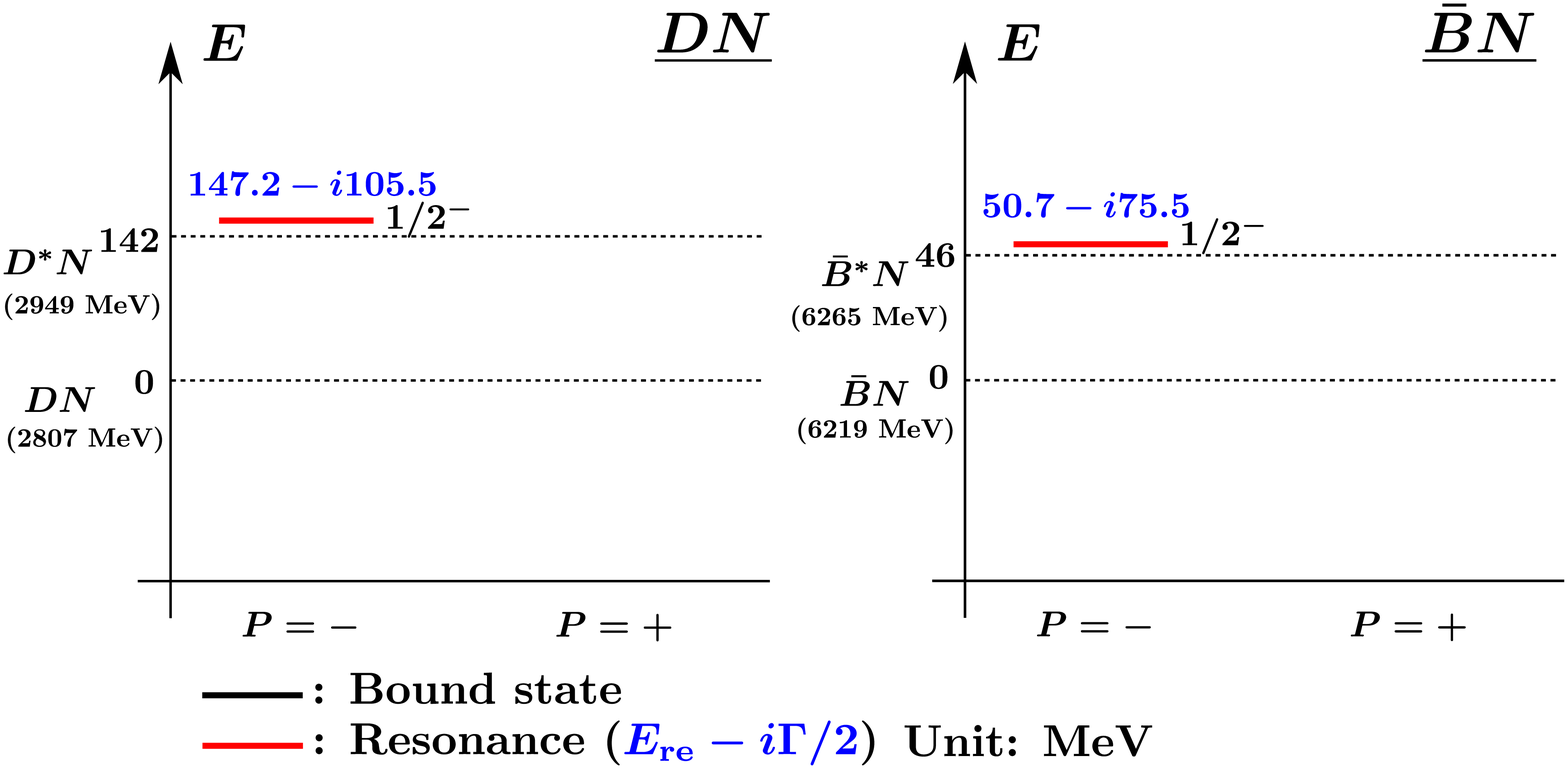}
 \caption{Resonances for $DN$ and $\bar{B}N$ with
 $I=1$.  The values are given when the $\pi\rho\,\omega$ potential is used.}
\label{Energy-LevelDNI=1}
\end{figure}
%%%%%%%%%%%%%%%%%%%%%%%%%%%%%%%%%%%%%%%%%%%%%%%%%%%%%%%%%%%%%%%%

For $I=1$ state, we found resonance only in $J^P=1/2^-$.
We show the energy level of the state in Fig.~\ref{Energy-LevelDNI=1}.
Small isospin factor suppresses a strong attraction as seen in
$\bar{D}N$ and $BN$ states.

%As seen in Figs.~\ref{Energy-LevelDN} and \ref{Energy-LevelDNI=1},
In non-exotic states in charm sector,
some charmed baryons are
already observed~\cite{Beringer:1900zz} or predicted by quark model
calculation~\cite{Copley:1979wj,Roberts:2007ni} near the $DN$ and $D^\ast N$ thresholds.
Such molecular states as displayed in Figs.~\ref{Energy-LevelDN} and
\ref{Energy-LevelDNI=1}
may have an effect on the structure of charmed baryon
states with same quantum number.

% many bound states and resonances of $DN$ and $\bar{B}N$ are found.
% Since some charmed baryons near the $DN$ and $D^\ast N$ thresholds are
% already observed or predicted by quark model
% calculation~\cite{Copley:1979wj,Roberts:2007ni},
% such molecular states may have an effect on the structure of charmed baryon
% states with same quantum number.

%==============================================
\section{Summary}
%==============================================
In the present study, we have investigated heavy baryons as hadronic
molecules formed by a heavy meson and a nucleon.
The $\pi$ exchange potential with heavy quark symmetry plays a
dominate role to compose bound states and resonances.
In particular, the tensor force of $\pi$ exchange is important to yield a strong
attractive force through channel couplings of different angular momentum
states of $\Delta L=2$.
On the other hand, the vector meson exchange plays a minor role in the
system.
The small mass splitting induces a strong $PN-P^\ast N$ mixing.
Therefore, states for bottom sector are more bound than states for
charm sector.

Many bound and resonant states have found in $I=0$ state.
However, there are few resonances with $I=1$ due to small isospin
factor.
Some loosely bound states and resonances are present in $\bar{D}N$ and
$BN$ exotic states.
For $DN$ and $\bar{B}N$ non-exotic states, the rich molecular structure 
may have an influence on the spectrum of 
the ordinary heavy baryon states.

\end{document}